\documentclass[final,3p,times,amssymb,qsymbols,twocolumn]{elsarticle}
 \biboptions{comma,sort&compress}
\pdfoutput=1
\usepackage{here}
\usepackage{graphicx,epstopdf}
 \usepackage{epsfig}

\def\nin{\noindent}

\def \Op{\mathcal{O} }

\newcommand{\bea}{\begin{eqnarray}}
\newcommand{\eea}{\end{eqnarray}}
\newcommand{\beq}{\begin{equation}}
\newcommand{\eeq}{\end{equation}}
\newcommand{\ec}{\end{center}}
\newcommand{\bc}{\begin{center}}

\newcommand{\pdir}{p\kern -5.2pt\raise 0.2ex\hbox {/}}

\newcommand{\vdir}{v\kern -5.75pt\raise 0.15ex\hbox {/}}
\newcommand{\kdir}{k\kern -5.75pt\raise 0.15ex\hbox {/}}
\newcommand{\epsdir}{\epsilon\kern -5.0pt\raise 0.15ex\hbox {/}}
\newcommand{\bvdir}{\bar{v}\kern -5.75pt\raise 0.15ex\hbox {/}}
\newcommand{\Ddir}{D\kern -7.75pt\raise 0.20ex\hbox {/}}
\newcommand{\Adir}{A\kern -7.75pt\raise 0.20ex\hbox {/}}
\newcommand{\ldir}{l\kern -5.0pt\raise 0.2ex\hbox{/}}
\newcommand{\varepsdir}{\varepsilon\kern -5.5pt\raise 0.15ex\hbox{/}}

\newcommand{\nn}{\nonumber}

\journal{Nuc. Phys. (Proc. Suppl.)}

\begin{document}

\begin{frontmatter}



\title{Complementarity of 
$B_s\to \mu^+\mu^-$ and  $B\to K \ell^+\ell^-$ in New Physics searches}

 \author[label1]{Damir Be\v cirevi\'c}
 \address[label1]{
Laboratoire de Physique Th\'eorique (B\^at.~210) and
Universit\'e Paris Sud, Centre d'Orsay, F-91405 Orsay Cedex, France.
} 
\author[label2]{Nejc Ko\v snik}
\address[label2]{LAL (B\^at. 200), Universit\'e Paris-Sud  B.P. 34,  91898 Orsay, France, and 
J. Stefan Institute, Jamova 39, P. O. Box 3000, 1001 Ljubljana, Slovenia
} 
 \author[label3]{Federico Mescia\corref{cor3}}
  \address[label3]{ECM \& ICCUB, Facultat de Fisica Universitat de Barcelona, Diagonal 647, E-08028 Barcelona, Spain.
}
 \author[label4]{Elia Schneider}
  \address[label4]{Dipartimento di Fisica, Universit\`a degli Studi di Trento,  and
INFN Gruppo collegato di Trento, 
Via Sommarive 14, Povo (Trento), I-38123 Italy.
}

\cortext[cor3]{Speaker}
\ead{mescia@ub.edu}



\begin{abstract}
\noindent
This year, LHC provided a very stringent bound on ${\rm Br}(B_s\to\mu^+\mu^-)$, bringing it closer to the value predicted by the Standard Model (SM). 
$B_s\to\mu^+\mu^-$ was believed to be the golden mode at LHCb to find SUSY because a large enhancement was expected in the regime of 
moderate and large values of $\tan\beta$. Other scenarios are still possible and a correlation with other decay channels is needed. We show that a 
complementary information on New Physics (NP) can be obtained model-indepedently from  the $B\to K \ell^+\ell^-$ decay mode. 
We provide a prediction ${\rm Br}(B\to K \ell^+\ell^-)$ based on the first lattice QCD results for all three relevant form factors, $f_{0,+}(q^2)$ and $f_{T}(q^2)$.
We were then able to provide the model-independent bounds on the complex couplings to scalar and pseudoscalr operators in the $b\to s$ sector.
\end{abstract}

\begin{keyword}
PACS: 13.25.Hw, 11.30.Er, 11.30.Hv.
\end{keyword}

\end{frontmatter}


\section{Introduction}
\nin

At Moriond 2012, LHCb lowered the upper
limit~\cite{LHCb-Bmumu} on ${\rm Br}(B_s\to\mu+\mu^-)$, pushing it very close to the Standard Model (SM)  value~\footnote{Recently, it has been noted that the effect of
  $B_s-\bar B_s$ mixing should be taken into account~\cite{Fleischer} and 
  whose net
  effect amounts to replacing $\mathrm{Br}(B_s \to \mu^+ \mu^-)^{\rm exp} 
  \to  (1-y_s) \mathrm{Br}(B_s \to \mu^+ \mu^-)^{\rm exp}$, where $y_s
  = (\Delta\Gamma/2\Gamma)_{B_s}
  = (9.0 \pm 2.1 \pm 0.8)\,\%$ was also measured by LHCb~\cite{1202.4717}. This
  correction is already incorporated in the corrected experimental 
  value~(\ref{bsmumu-exp}).} 
\bea\label{bsmumu-exp}
{\rm Br} \left(B_{s} \to \mu^+ \mu^- \right)^{\rm exp} <    4.1 \times 10^{-9}
\,.
\eea
The current upper bound is now 
only about $1.3$ times larger than the SM one, namely
\bea
{\rm Br} \left(B_{s} \to \mu^+ \mu^- \right)^{\rm th-SM} =  (3.3 \pm 0.3)\times 10^{-9}.
\eea
In $2011$, before the LHC results appeared, the best bound was from CDF~\cite{Aaltonen:2011fi} and it was about $10$ times the SM value.

The fact that the current bound on ${\rm Br} \left(B_{s} \to \mu^+ \mu^- \right)$ is closer to the SM forces us
to scrutinize  this process and disentangle the possible NP effects. To that end, other processes such as $B\to K \ell^+\ell^-$
become particularly helpful~\cite{nostro}.

Despite its closeness to the SM prediction, $B_s\to\mu+\mu^-$ 
remains interesting. The SM contribution to this decay is helicity suppressed whereas the contribution of the 
(pseudo)scalar operators are not. In SUSY the extended Higgs sector provides a natural scalar contributions and enhances significantly the  ${\rm Br} \left(B_{s} \to \mu^+ \mu^- \right)$.

In a model-independent scenario the (pseudo) scalar couplings are free complex parameters and 
can enhance or suppress ${\rm Br} \left(B_{s} \to \mu^+ \mu^- \right)$. Importantly, however, the same (pseudo) scalar couplings affect the $B \to K\mu^+ \mu^-$ decay in a complementary way, namely
\begin{eqnarray}
{\rm Br} \left(B_{s} \to \mu^+ \mu^- \right) &\propto& f\left[
m_\mu^2 \left(C_{10}-C_{10}^\prime\right),\right.\nn\\
&&\quad\left.\left(C_{P,S}-C_{P,S}^\prime\right)\right]\nn\\
{\rm Br} \left(B \to K \ell^+ \ell^- \right) &\propto& g\left[
\left(C_{7,9,10}+C_{7,9,10}^\prime\right),\right.\nn\\
&&\quad\left.\left(C_{P,S}+C_{P,S}^\prime\right),C_T,C_{T5}\right],\nn 
\eea
and the two can be used to unravel the couplings  $C_i$ which encode
the short-distance physics information that, at $m_b$ scale, enter the $b\to s$ Hamiltonian as (see \cite{nostro} for details),
\begin{eqnarray} \label{eq:Heff}
 H_{\rm{eff}} = \sum_{i=7,8,9,10,P,S,T,T5} \left(C_i  \mathcal O_i + C'_i  \mathcal O'_i\right) \,,
\end{eqnarray}
The operator basis in which the Wilson coefficients $C_i$ have been computed in the SM is:
\bea\label{basisOps}
{\mathcal{O}}_{7}^{(\prime)} &=& \frac{e}{g^2} m_b
(\bar{s} \sigma_{\mu \nu} P_{R,L} b) F^{\mu \nu} ,\nn \\
{\mathcal{O}}_{9}^{(\prime)}  &=& \frac{e^2}{g^2} 
(\bar{s} \gamma_{\mu} P_{L,R} b)(\bar{\ell} \gamma^\mu \ell) ,\nn \\
{\mathcal{O}}_{10}^{(\prime)} &=&\frac{e^2}{g^2}
(\bar{s}  \gamma_{\mu} P_{L,R} b)(  \bar{\ell} \gamma^\mu \gamma_5 \ell) ,\nn \\
{\mathcal{O}}^{(\prime)}_{S} &=&\frac{e^2}{16\pi^2}
 (\bar{s} P_{R,L} b)(  \bar{\ell} \ell) , \nn \\
 {\mathcal{O}}_{P}^{(\prime)} &=&\frac{e^2}{16\pi^2}
 (\bar{s} P_{R,L} b)(  \bar{\ell} \gamma_5 \ell) ,\nn \\
 {\mathcal{O}}_{T} &=&\frac{e^2}{16\pi^2}
 (\bar{s} \sigma_{\mu\nu} b)(  \bar{\ell}\sigma^{\mu\nu} \ell) ,\nn \\
 {\mathcal{O}}_{T5} &=&\frac{e^2}{16\pi^2}
 (\bar{s} \sigma_{\mu\nu} b)(  \bar{\ell} \sigma^{\mu\nu}\gamma_5 \ell),
\eea
$C_{7,9,10}$ receive contributions in SM from the  $W$- and $Z$- boxes 
and penguin diagrams, while $C_{P,S}^{(\prime)}$, $C_{7,9,10}^\prime$  and $C_{T,T5}$  are totally negligible in the SM.

At Moriond 2012, the results of the first accurate measurement of ${\rm Br} \left(B \to K \ell^+ \ell^- \right)$ 
were presented by BaBar~\cite{exp-BKll}, 
\bea\label{exp-Kll}
{\rm Br} \left(B \to K \ell^+ \ell^- \right) = \left(4.7\pm 0.6\right)\times 10^{-7}.
\eea
Within their statistics BaBar observe no isospin asymmetries between $B^+\to K^+\ell^+ \ell^-$ and $B^0_d\to K^0 \ell^+ \ell^-$ and make the average 
of the two modes assuming the lepton flavor universality. LHCb, very recently, presented their results and quoted~\cite{LHC-BKll}
\bea
{\rm Br} \left(B^0_d \to K^0 \mu^+ \mu^- \right) = \left(3.1^{+0.07}_{-0.06}\right)\times 10^{-7}.
\eea
LHCb and BaBar results agree within one standard deviation.  A surprising feature of the LHCb result is that, contrary to BaBar, they observe a large 
isospin asymmetry, a puzzling phenomenon that could partly be described by the effect of (structure dependent) soft photons. That issue is beyond the scope of this paper and will be addressed elsewhere. After comparing the above experimental results with our theory estimate (see next section)
\bea\label{eq:99}
{\rm Br} \left(B \to K \ell^+ \ell^- \right)^{SM} = \left(7.0\pm 1.8\right)\times 10^{-7},
\eea
we see that the BaBar result agrees well with theory, whereas the LHCb is by more than one-sigma lower. 
In the following we first discuss the theoretical (hadronic) uncertainties entering the
${\rm Br} \left(B_s\to \mu^+ \mu^- \right)$ and ${\rm Br} \left(B \to K \ell^+ \ell^- \right)$, and then discuss the model independent information about NP that can be deduced from a simultaneous study of these two decay modes.   

\section{Hadronic uncertainties in $B_{s} \to \mu^+ \mu^- $ and $B \to K\mu^+ \mu^-$}
\nin
The SM prediction of $B_{s} \to \mu^+ \mu^- $ has been recently revisited in~\cite{Buras:2012ru}, where one part of the soft photons has been included. The main QCD uncertainty in
\bea\label{eq:88}
\!\!\!\!\!\!\!\!\!\!\!\!\!{\rm Br} \left( B_s \to \mu^+ \mu^- \right)\!\!\!\!& \propto&\!\!\!\!\!\! f_{B_s}^2 \left[ 
\Big| C_S -C_S^\prime \Big|^2 \right. \\
& +&\!\!\!\!\!\! \left.\Big|  \left( C_P -C_P^\prime \right) +  { 2  m_\mu m_b \over m_{B_s}^2} \left( C_{10} - C_{10}^{\prime  } \right) \Big|^2 \, \right] \, ,
\nn\eea
comes from the decay constant $f_{B_s}$, defined via,
\bea
\langle 0 | \bar b\gamma^\mu\gamma_5 s| B^0_s(p)\rangle =i p^\mu f_{B_s}\,,
\eea 
\begin{figure}
\begin{center}
\includegraphics[width=7.8cm]{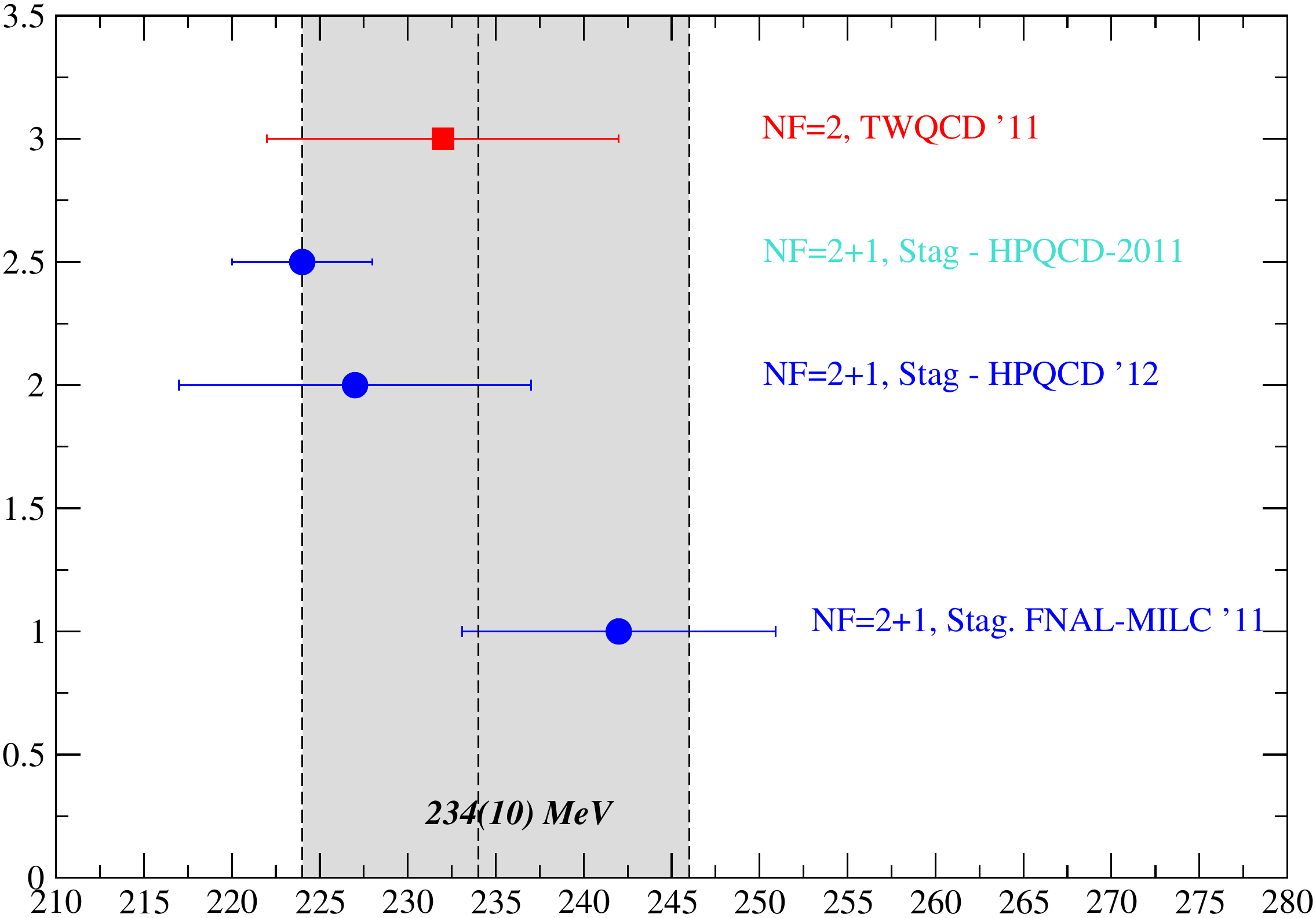}
\caption{\label{fig:1}\footnotesize{\sl 
Recent lattice results for $f_{B_s}$.}} 
\end{center}
\end{figure}
that has been computed by means of lattice QCD (with $N_f=2$ and $2+1$ light flavors) by many groups. Recent results obtained in full QCD are shown in fig.~\ref{fig:1}, 
from which we estimate $f_{B_s}=(234\pm 10)$ MeV. Note that the resulting $\sim 4\%$ uncertainty used to be $20\%$ and its reduction is a net result of recent progress in lattice QCD~\cite{cecilia}. The uncertainty in $f_{B_s}$ implies an overall  $\sim 7\%$ theory error on ${\rm Br} \left(B_s\to \mu^+ \mu^- \right)$. 

As far as ${\rm Br} \left(B \to K \ell^+ \ell^- \right)$ is concerned, the dominant uncertainties are those associated with the following two hadronic matrix elements, 
\begin{center}
\bea
\langle K | \bar b \gamma^\mu s| B^0_s\rangle \propto  f_{+,0}(q^2)\,, \nn\\
\langle K | \bar b \sigma^{\mu\nu} s| B^0_s\rangle\propto f_{T}(q^2)\,, 
\eea
\end{center}
parameterized by the three form factors $f_{+,0,T}(q^2)$, which are either computed in the numerical simulations of quenched QCD on the lattice (LQCD), or by the QCD sum rule analysis near the light cone (LCSR)~\cite{ball-zwicky}. 
\begin{figure}
\begin{center}
\includegraphics[width=7.3cm]{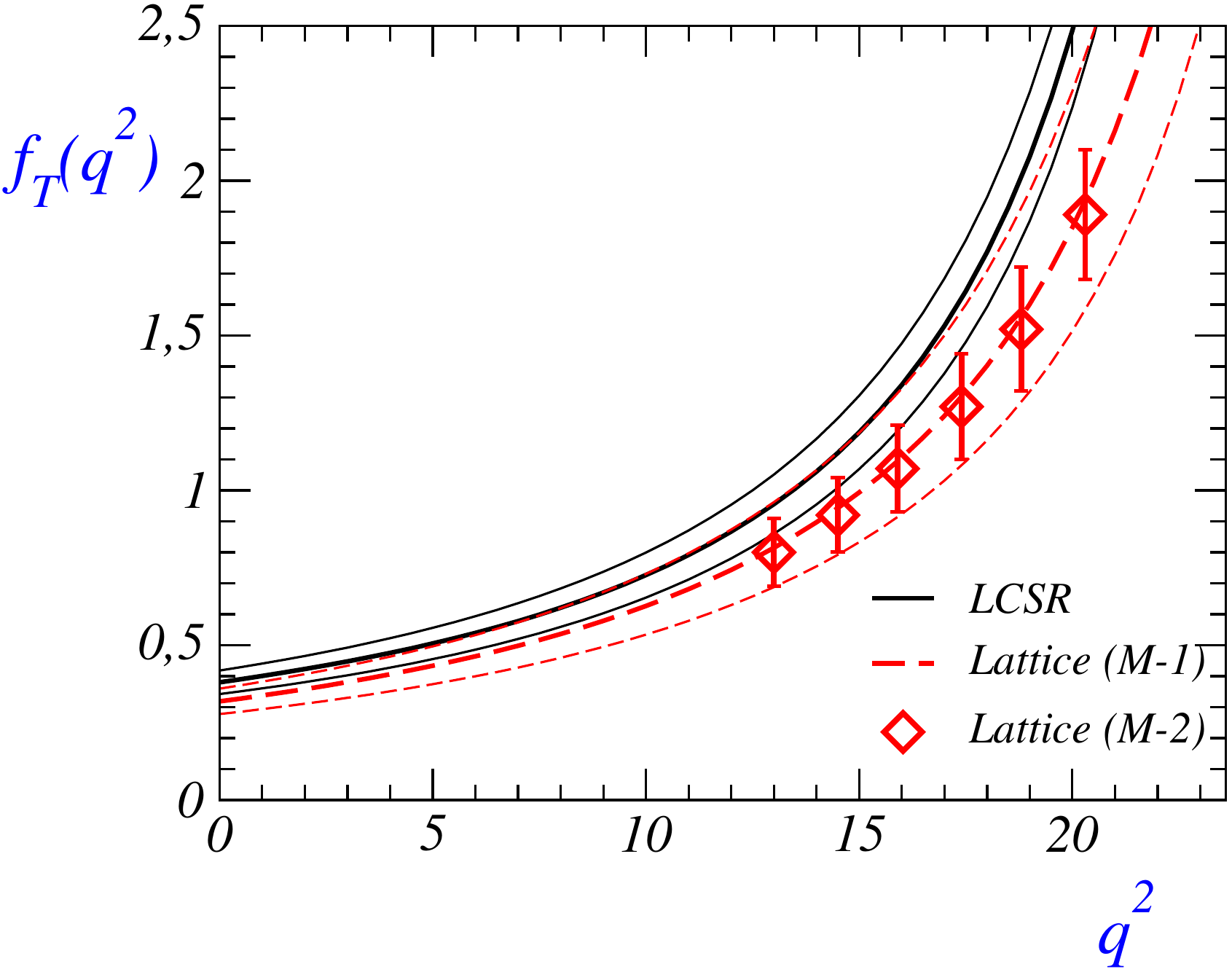}
\caption{\label{fig:11}\footnotesize{\sl 
Results of the first (quenched) lattice QCD determination of the form factor $f_T(q^2)$ are plotted together with the results obtained by using LCSR.}} 
\end{center}
\end{figure}
Since we present the first lattice QCD determination of the $f_{T}(q^2)$ form factor, we plot its shape and values in fig.~\ref{fig:11} along with the estimates made by using LCSR.
LQCD results appear to be consistent with those obtained from LCSR within $30\%$ of error in results obtained by both methods. This leaves room 
for improvement by the new generation of unquenched LQCD simulations. Several such studies are underway~\cite{bkk-latt}. 
Notice that the improvement of $f_{+,0,T}(q^2)$ is much more realistic to expect than those parameterizing the $B\to K^\ast$ transition matrix elements, because the latter decay involves many more form factors, with at least three of them being subject to very large uncertainties (see e.g. ref.~\cite{our-vec}). 
With the currently available estimates of the form factors we obtain~\cite{bobeth}
\bea\label{Kll-sm} 
\!\!\!\!\!\!\!\!
{\rm Br} \left(B \to K\ell^+ \ell^- \right)_{\rm SM}  = \left\{ \begin{array}{ll}
         (7.5 \pm 1.4 )\times 10^{-7} & \mbox{\rm LQCD}\nn\\
         &\\
         (6.8 \pm 1.6 )\times 10^{-7} & \mbox{\rm LCSR}
\end{array} \right. . \nn
\eea 
The result quoted in eq.~(\ref{eq:99}) covers both of the above values. 
For our purpose it is crucial to note that contrary to eq.~(\ref{eq:88}), the expression for ${\rm Br} \left(B \to K\ell^+ \ell^- \right)$ involves the sums of the Wilson coefficients $C_{10,S,P}+
C_{10,S,P}^\prime$, and therefore the two decays provide the complementary information about NP. A detailed analysis of various such scenarios is presented in ref.~\cite{nostro}. Here we focus on one example.
\section{A New Physics Scenario}

As an example of the  complementarity of constraints we add to the SM 
a scalar and a pseudoscalar operators,  
\bea
H^{\rm NP} = C_S \Op_S + C_P\Op_P \ .
\eea
derive the expressions for ${\rm Br} \left(B_s\to \mu^+\mu^- \right)^{\rm NP}$ and ${\rm Br} \left(B \to K\ell^+ \ell^- \right)^{\rm NP}$, and then plot the allowed values for $|C_{S,P}|$  
consistent with eqs.(\ref{bsmumu-exp}) and (\ref{exp-Kll}), respectively.
\begin{figure*}[t!]
\begin{center}
{\resizebox{5.1cm}{!}{\includegraphics{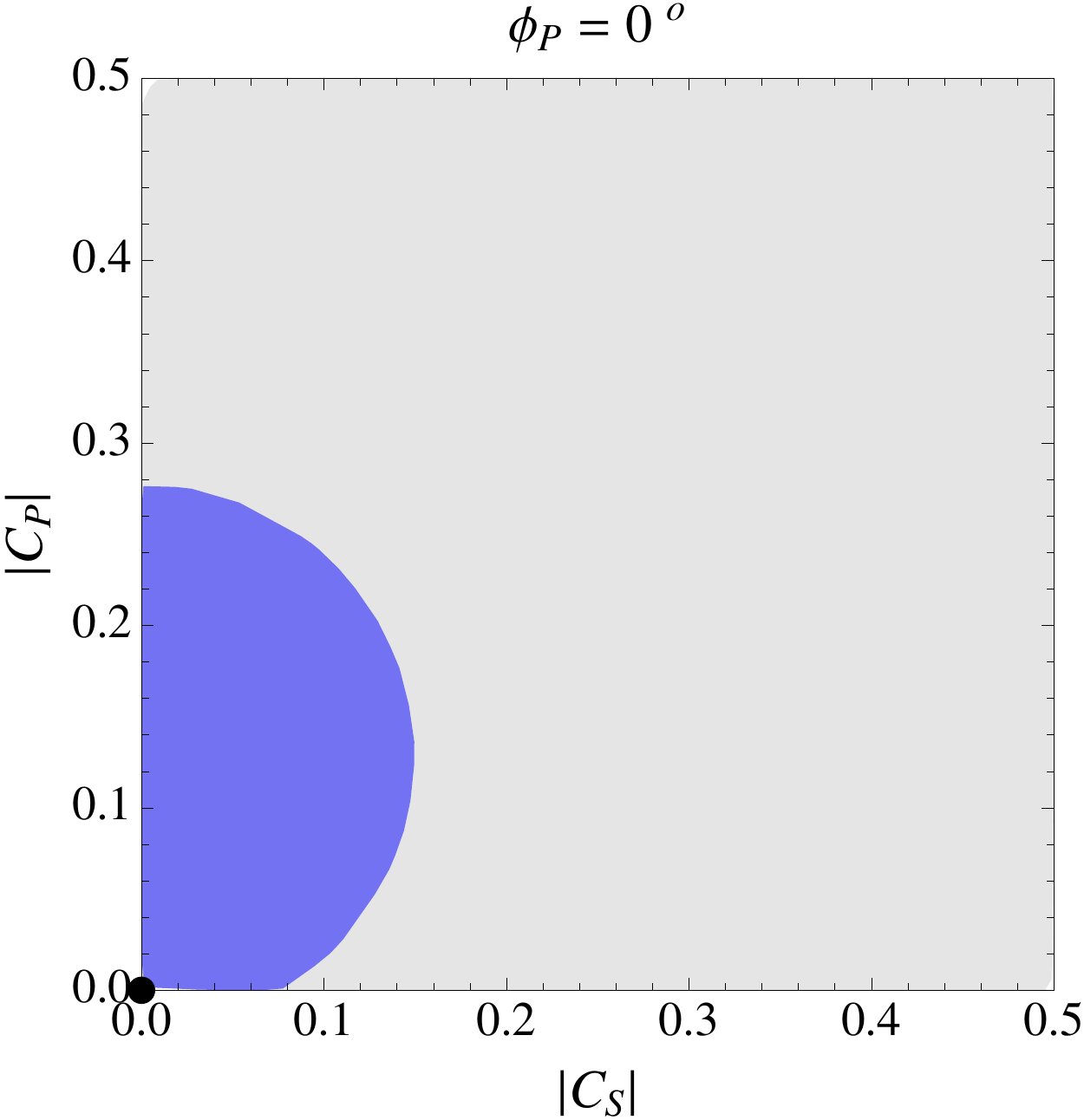}}} 
{\resizebox{5.1cm}{!}{\includegraphics{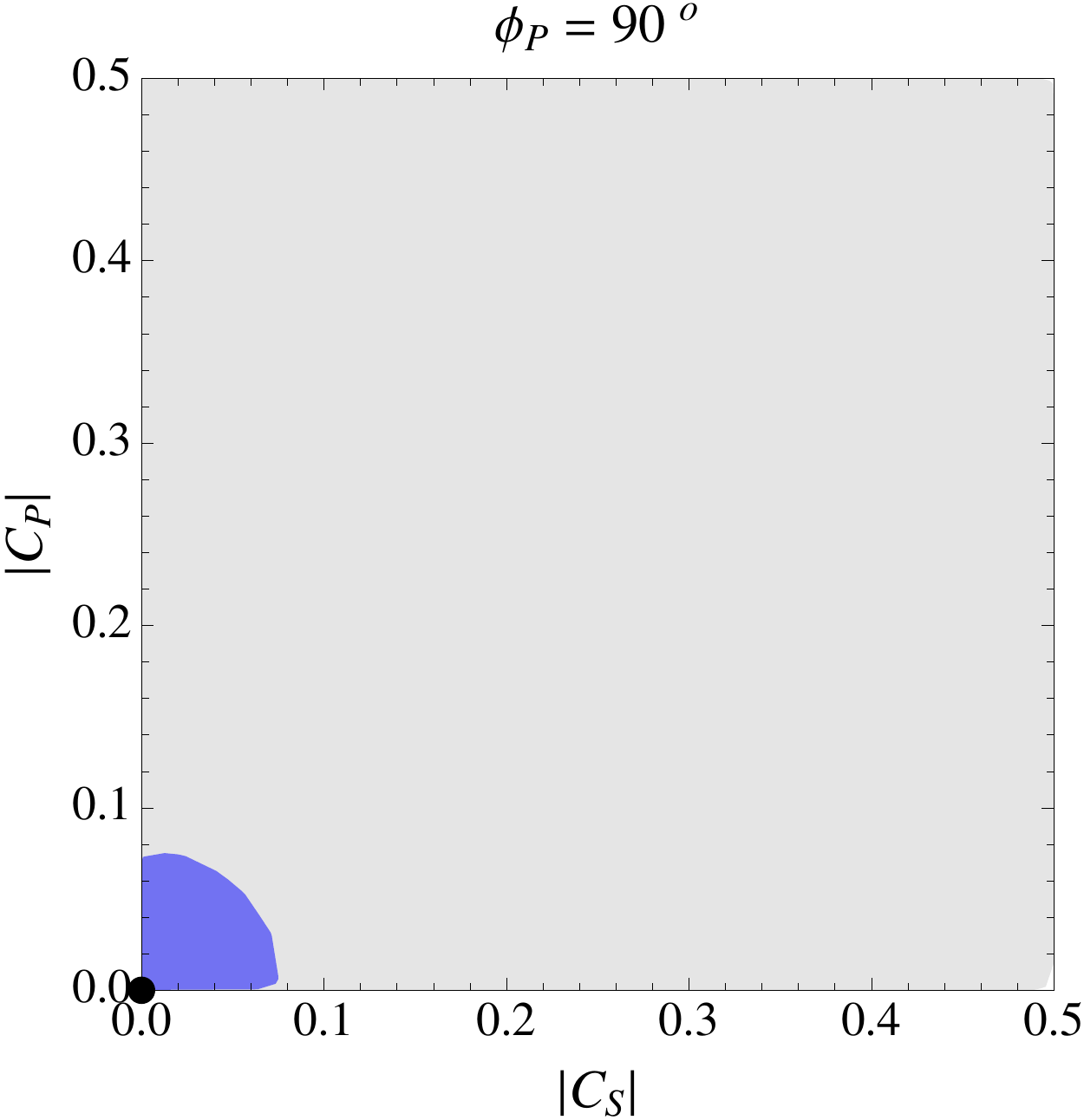}}} 
{\resizebox{5.1cm}{!}{\includegraphics{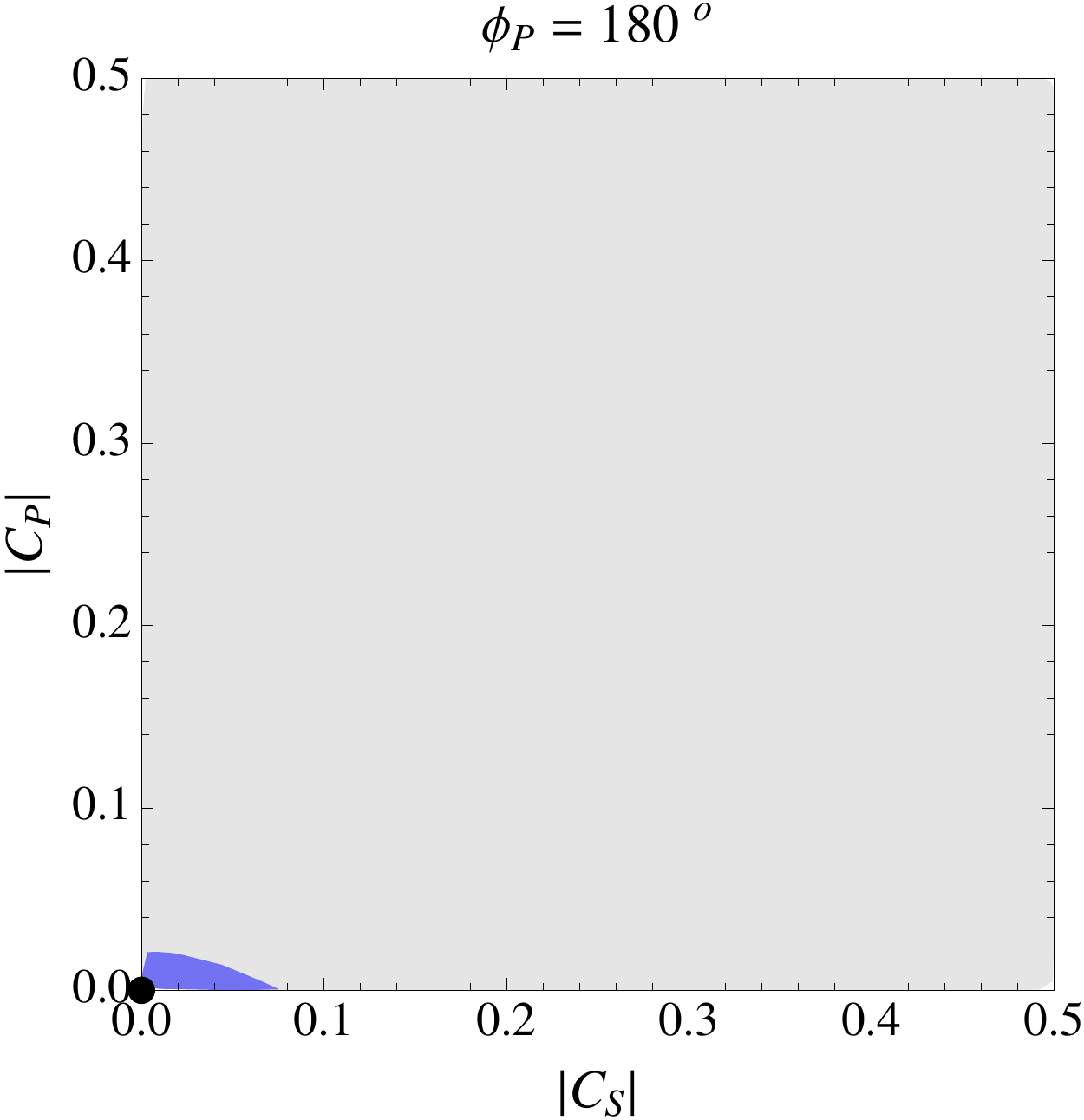}}} 
\caption{\label{fig:A}\footnotesize{\sl 
Allowed values for $ \left| C_{S} \right|$  and $ \left|  C_{P} \right|$ obtained by combining  the experimental information on  ${\rm Br} \left(B \to K\ell^+ \ell^- \right)$ (light shaded area) and the upper bound on  ${\rm Br} \left(B_s \to \mu^+ \mu^- \right)$ (dark shaded area). Illustration is provided for three various values of the relative phase $\phi_P$. }} 
\end{center}
\end{figure*}
The largest range of $|C_{S,P}|\neq 0$ is obtained when
the pseudoscalar coupling is real, $\phi_P=0$. 
Considering the case $\phi_P\neq 0$ is essential because $C_{P}$ interferes with 
$C_{10}$. Note that the negative interference could lead to ${\rm Br} \left(B_s \to \mu^+ \mu^- \right)$ even smaller than the one predicted in the SM [c.f. eq.~(\ref{eq:88})]. $C_{S}$, instead, only enters via its moduli. 

We observe that at present the current constraint
provided by $B\to K\ell^+\ell^-$ is redundant for this particular scenario, but that situation could
radically change if the errors on $B\to K$ form factors were significantly
reduced. If we keep the central values of the form
factors fixed and reduce the errors by $20\%$ the measured and theoretically evaluated ${\rm Br}
\left(B \to K\ell^+ \ell^- \right)^{\rm SM}$ would not be compatible and $B \to K\ell^+ \ell^- $ 
would become the essential constraint on the possible values of $|C_{S,P}|$. 
In fig.~\ref{fig:9} we show the cases for which the overlapping region (satisfied by both constraints) exists. E.g. for $\phi_P\gtrsim
40^\circ$ such a solution would not exist, which would be a valuable
information about NP.
\begin{figure*}[th!]
\begin{center}
{\resizebox{5.1cm}{!}{\includegraphics{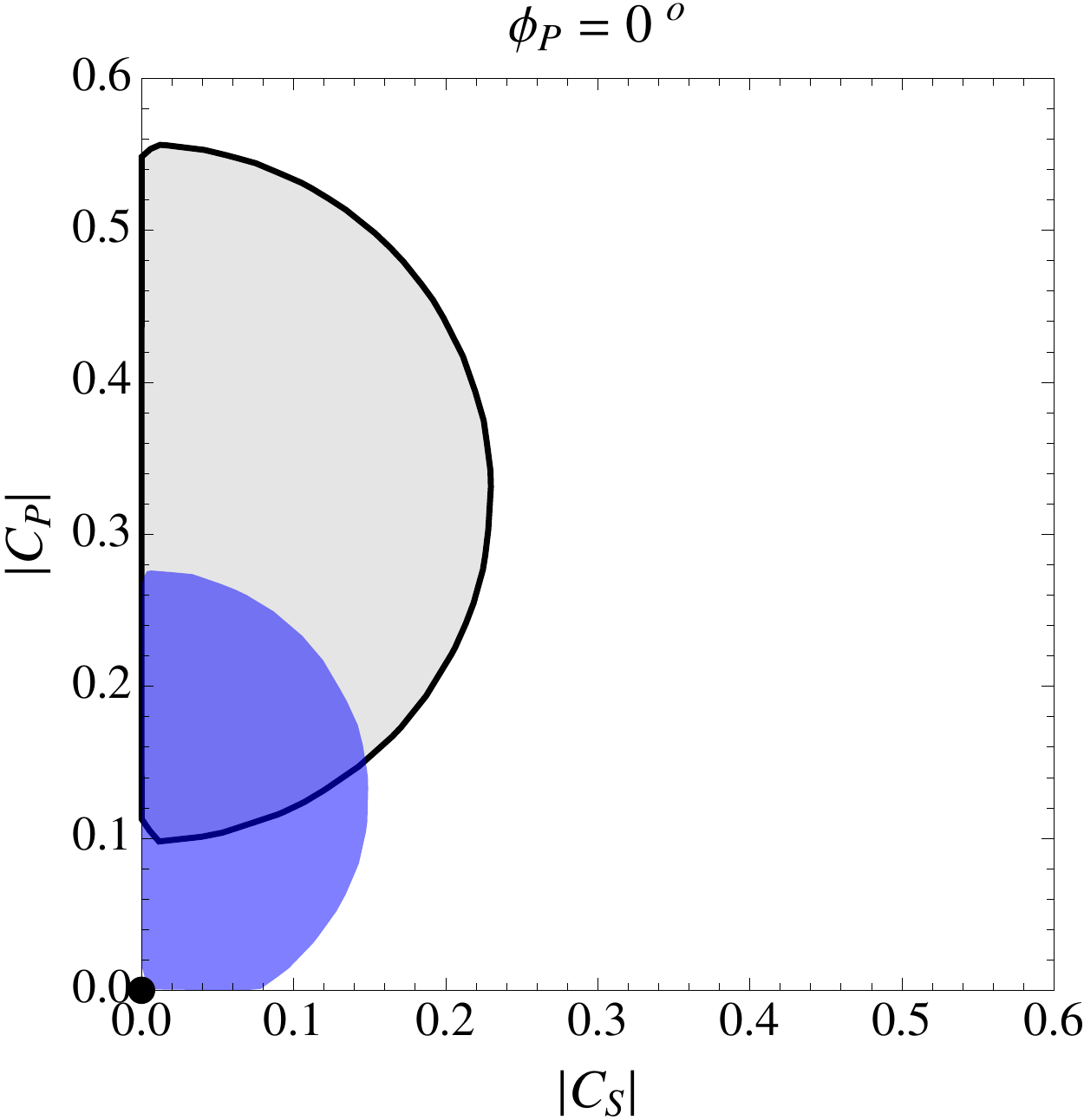}}} 
{\resizebox{5.1cm}{!}{\includegraphics{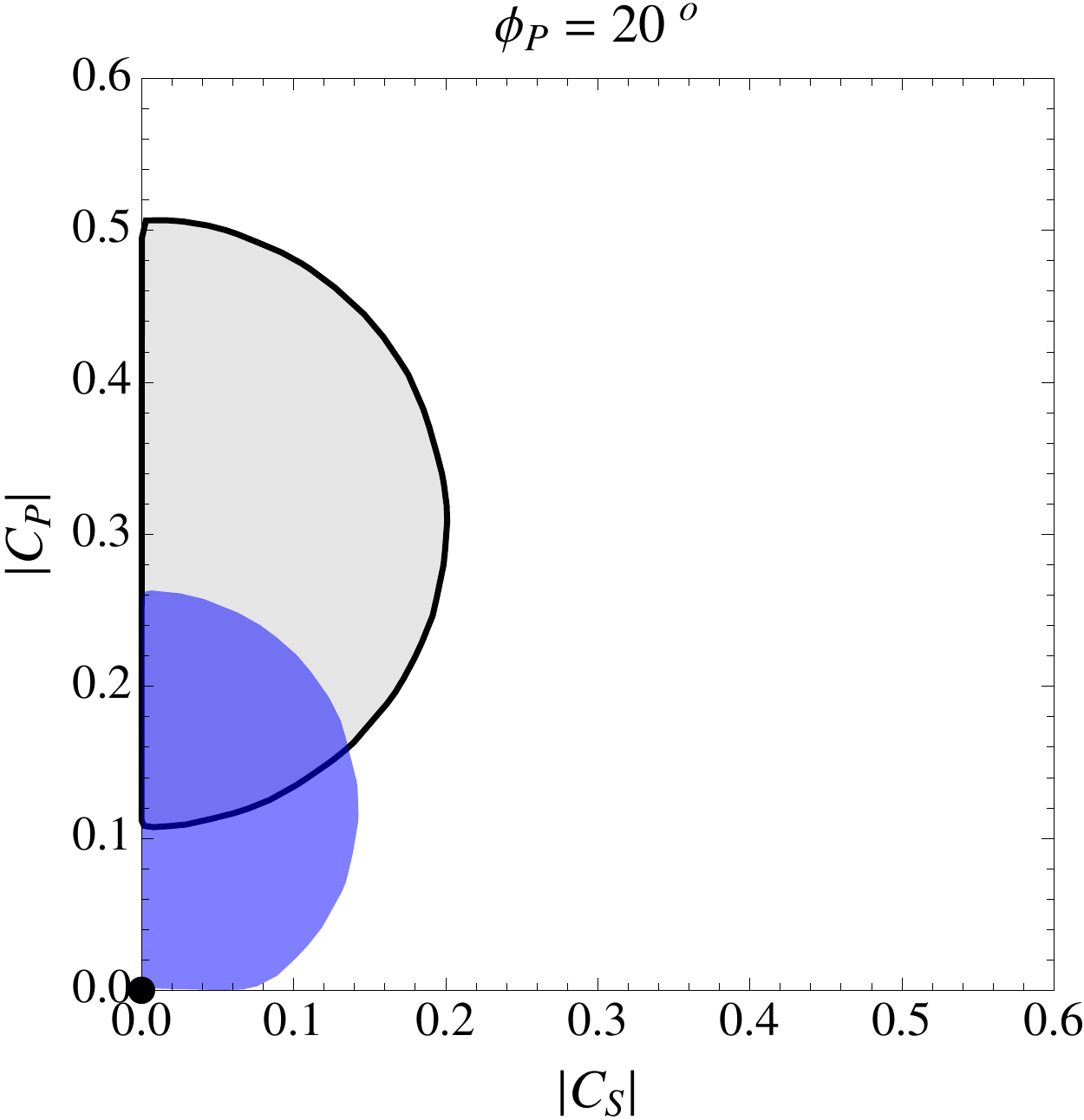}}} 
{\resizebox{5.1cm}{!}{\includegraphics{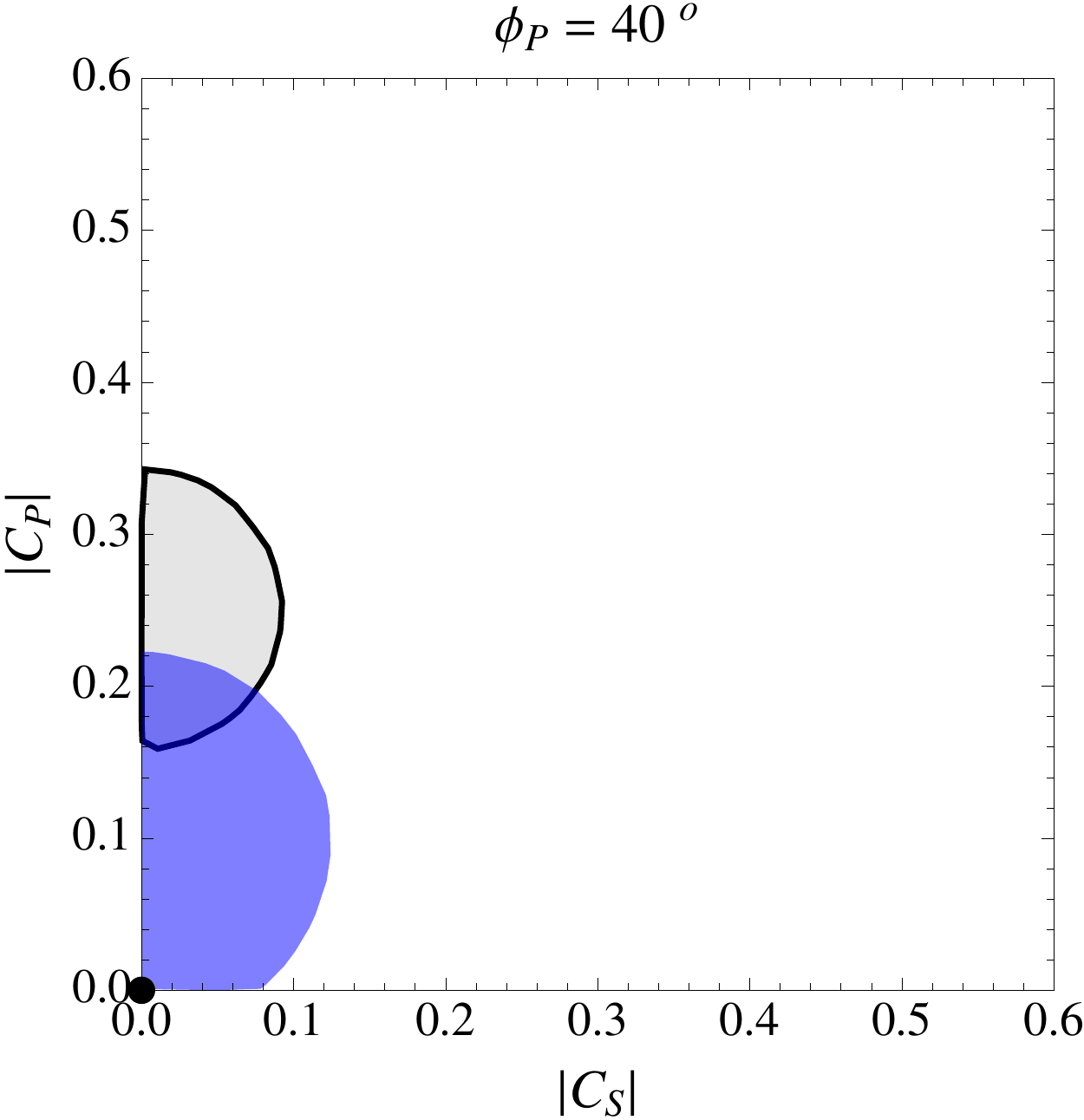}}} 
\caption{\label{fig:9}\footnotesize{\sl 
Same as in fig.~\ref{fig:A}, except that the errors on the hadronic form factors relevant to $B \to K\ell^+ \ell^- $ are reduced by $20\%$, and we plot the cases with $\phi_P=0$, $20^\circ$ and $40^\circ$. }} 
\end{center}
\end{figure*}
This further exacerbates the necessity for a better control of errors
on the $B \to K$ form factors, $f_{+,0,T}(q^2)$. Furthermore, the experiment effort to measure
the partial decay width of $B \to K\ell^+\ell^- $ at large values of $q^2$ (see ref.~\cite{exp-BKll})  would be highly
welcomed because that range of $q^2$'s is more convenient for the precision computations in LQCD.

\section{Conclusions}
${\rm Br} \left(B_s \to K \mu^+ \mu^- \right)$ is sensitive to (pseudo)scalar operators, $O_{S}^{(\prime)}$ and $O_{P}^{(\prime)}$,
which are to large extent absent in the SM. Importantly, only one hadronic parameter needs to be controlled, $f_{B_s}$, which nowadays has only 
small theoretical uncertainties.

${\rm Br} \left(B \to K \ell^+ \ell^- \right)$ is very sensitive 
to (pseudo)scalar operators too, and the theoretical predictions involve 3 form factors, which have been computed both in LQCD and by using the LCSR, but still suffer from large errors that can be substantially reduced if computed in the new generation of full LQCD simulations.  Importantly, this decay mode offers a complementary information to the coupling to NP particles, with respect to the one that can be deduced from ${\rm Br} \left(B_{s} \to \mu^+ \mu^- \right)$.

These modes constitute the essential probe to NP couplings that could be combined with the other theoretically clean observables, such as the transverse asymmetries in $B\to K^* \ell^+ \ell^-$~\cite{quim}.

\section*{Acknowledgements}
\nin
F.M. thanks the LPTA-Montpellier for hospitality and acknowledges financial support 
from FPA2010-20807 and the Consolider
CPAN project.




\begin{thebibliography}{999}
\vspace*{-0.25cm}
\bibitem{LHCb-Bmumu} 
  R.~Aaij {\it et al.}  [LHCb Collaboration],
  arXiv:1203.4493.
  \bibitem{Fleischer}
  K.~De Bruyn {\it et al.} ,
  Phys.\ Rev.\ Lett.\  {\bf 109} (2012) 041801;
  S.~Descotes-Genon,{\it et al.} ,
  Phys.\ Rev.\ D {\bf 85}, 034010 (2012)

\bibitem{1202.4717}
  R.~Aaij {\it et al.}  [LHCb Collaboration],
  Phys.\ Rev.\ Lett.\  {\bf 108} (2012) 101803,
  {\it ibid} 241801.

\bibitem{Aaltonen:2011fi} 
  T.~Aaltonen {\it et al.}  [CDF Collaboration],
  Phys.\ Rev.\ Lett.\  {\bf 107}, 239903 (2011)

\bibitem{exp-BKll} 
J.P.~Lees et al.  [BABAR Collaboration],
  arXiv:1204.3933.

\bibitem{LHC-BKll}
  R.~Aaij {\it et al.}  [LHCb Collaboration],
  JHEP {\bf 1207} (2012) 133.


\bibitem{nostro} 
  D.~Becirevic, {\it et al.},
  arXiv:1205.5811 [hep-ph].

\bibitem{Buras:2012ru} 
  A.~J.~Buras  {\it et al.},
  arXiv:1208.0934 [hep-ph].

\bibitem{cecilia} 
  V.~Lubicz and C.~Tarantino,
  Nuovo Cim.\ B {\bf 123}, 674 (2008).

\bibitem{ball-zwicky}
  P.~Ball and R.~Zwicky,
  Phys.\ Rev.\  D {\bf 71} (2005) 014029;
A.~Khodjamirian  {\it et al.},
  Phys.\ Rev.\ D {\bf 75} (2007) 054013.

\bibitem{bkk-latt}
R.~Zhou {\it et al.}  [Fermilab Lattice and MILC Collaborations],
  arXiv:1111.0981 [hep-lat];
  D.~Becirevic  {\it et al.}, in preparation.


\bibitem{our-vec}
  A.~Abada, {\it et al.}
                  [SPQcdR collaboration],
  Nucl.\ Phys.\ Proc.\ Suppl.\  {\bf 119} (2003) 625;
  K.~C.~Bowler, {\it et al.}  [UKQCD
                  Collaboration],
  JHEP {\bf 0405} (2004) 035;
 D.~Becirevic, {\it et al.}
  Nucl.\ Phys.\ B {\bf 769}, 31 (2007).


\bibitem{bobeth}
  C.~Bobeth,  {\it et al.},
  JHEP {\bf 0712} (2007) 040.

\bibitem{quim}
  F.~Kruger and J.~Matias,
  Phys.\ Rev.\  D {\bf 71} (2005) 094009;
U.~Egede, {\it et al.},
  JHEP\ {\bf 0811}, 032  (2008).
D.~Becirevic and E.~Schneider,
  Nucl.\ Phys.\ B {\bf 854} (2012) 321;
  JHEP {\bf 1204}, 104 (2012);
S.~Descotes-Genon, {\it et al.},
  arXiv:1207.2753 [hep-ph].
C.~Bobeth {\it et al.},
  JHEP\ {\bf 1007}, 098  (2010).


\end{thebibliography}
\end{document}